\documentclass[twocolumn,preprintnumbers,secnumarabic,amsmath,amssymb,superscriptaddress,nofootinbib]{revtex4}

\usepackage{graphicx}
\usepackage{xcolor}
\usepackage{dcolumn}
\usepackage{bm}

\newcommand{\beq}{\begin{equation}}
\newcommand{\eeq}{\end{equation}}

\newcommand{\bea}{\begin{eqnarray}}
\newcommand{\eea}{\end{eqnarray}}

\def\aD{\overline{\mathrm{D3}}}

\begin{document}

\preprint{CPHT-RR008.022022, CERN-TH-2022-013}
\vspace*{-.1cm}
\title{Bare-Bones de Sitter}

\smallskip
\author{Iosif Bena}

\affiliation{{Institut de Physique Th\'eorique,
	Universit\'e Paris Saclay, CEA, CNRS,
 F-91191 Gif-sur-Yvette, France
}}

\author{Emilian Dudas}
\affiliation{{Centre de Physique Th\'eorique, CNRS, Ecole Polytechnique, IP Paris, F-91128 Palaiseau, France }}
\affiliation{{CERN, Theory Division, Geneva, Switzerland}}

\author{Mariana Gra\~na}

\affiliation{{Institut de Physique Th\'eorique,
	Universit\'e Paris Saclay, CEA, CNRS,
 F-91191 Gif-sur-Yvette, France
}}

\author{Gabriele Lo Monaco}

\affiliation{{Institut de Physique Th\'eorique,
	Universit\'e Paris Saclay, CEA, CNRS,
 F-91191 Gif-sur-Yvette, France
}}

\author{Dimitrios Toulikas}

\affiliation{{Institut de Physique Th\'eorique,
	Universit\'e Paris Saclay, CEA, CNRS,
 F-91191 Gif-sur-Yvette, France
}}

\preprint{}

\begin{abstract}

We compute the supersymmetry-breaking three-form fluxes generated by the addition of anti-D3 branes at the tip of a Klebanov-Strassler throat. These fluxes give rise to nontrivial terms in the superpotential when the throat is embedded in a flux compactification. We describe these terms both from a ten-dimensional and from a four-dimensional perspective and show that, upon including K\"ahler-moduli stabilization, the resulting potential admits de Sitter minima. Our proposed de Sitter construction does not require additional supersymmetry-breaking  $(0,3)$ fluxes, and hence is more minimalist than the KKLT proposal.

\end{abstract}

\maketitle

\section{Introduction}
\vspace*{-.3cm}
The accelerated expansion of our Universe points towards the existence of a positive vacuum energy density, whose value is about 120 orders of magnitudes smaller than the value expected from field-theory estimates. On the other hand, there are by now several arguments \cite{Obied:2018sgi} that stable de Sitter vacua cannot be constructed in controlled low-energy effective theories that are consistent with quantum gravity. This leaves only two open possibilities: either the accelerated expansion of our Universe comes from a time-dependent vacuum energy density, or there is a problem with the no-de-Sitter conjecture, which can be disproved by an explicit construction.

Unfortunately, constructing metastable de Sitter vacua is notoriously difficult in String Theory.  Despite its intricate ingredients, short-comings and potential instabilities, the almost-twenty-year-old construction of Kachru, Kallosh, Linde and Trivedi (KKLT) \cite{Kachru:2003aw} still stands out as one of the very few generic proposals that has not been fully proven to be unstable. It is a three-step construction that combines fluxes, non-perturbative phenomena and anti-D3 branes in a warped Calabi-Yau compactification with a deformed conifold-type throat. In order to obtain a positive and small cosmological constant, the fluxes required in the first step need to break supersymmetry generating a very small superpotential $W_{0,\text{KKLT}}$. This has been criticized on two counts: theory and practice. On the formal side, these supersymmetry-breaking runaway solutions are not protected against corrections, and it was argued in \cite{Sethi:2017phn} that they are not a good ground onto which one can add the non-perturbative ingredients necessary in the second step to prevent the runaway and stabilize the volume moduli. On the practical side, it is very hard to obtain explicit solutions with a sufficiently small superpotential, although there has been recent progress in engineering this type of flux vacua \cite{Demirtas:2019sip,Demirtas:2020ffz,Bastian:2021hpc}. 

The purpose of this Letter is to take a step towards bridging the conflict between the no-de-Sitter swampland arguments \cite{Obied:2018sgi} and what can be constructed explicitly and controllably in String Theory. We propose a new method to construct de Sitter vacua, which has one less ingredient than the KKLT construction, and hence is potentially plagued by less problems. More precisely, we show that one can construct de Sitter vacua with a small 
cosmological constant without the need of a flux superpotential $W_{0,\text{KKLT}}$.

Our key observation is that the anti-D3-branes necessary to uplift the cosmological constant source fluxes that generate precisely a small superpotential. Therefore, in our ``bare bones" de Sitter construction, only supersymmetric fluxes are needed in the first step, thus avoiding the problems mentioned above. 
\vspace*{-.5cm}
\section{Fluxes generated by  $\overline{\mathrm{D3}}$ branes}
\label{sec:fluxes}
\vspace*{-.3cm}

A strongly warped region in a Calabi-Yau compactification can be engineered as a Klebanov-Strassler (KS) throat \cite{Klebanov:2000hb}. This is a cone over an $S^2\times S^3$ base (see Figure 1). The two-sphere of the base shrinks at the tip of the cone while the three-sphere has always finite size, parameterized by a modulus $Z$. The base can be also thought as a $U(1)$ fibration over $S^2\times S^2$. The symmetries of the geometry consist of two $SU(2)$ factors  acting on the base two-spheres and a $\mathbb{Z}_2$ swapping them.
\begin{figure}[!ht]
\centering \includegraphics[scale=0.6]{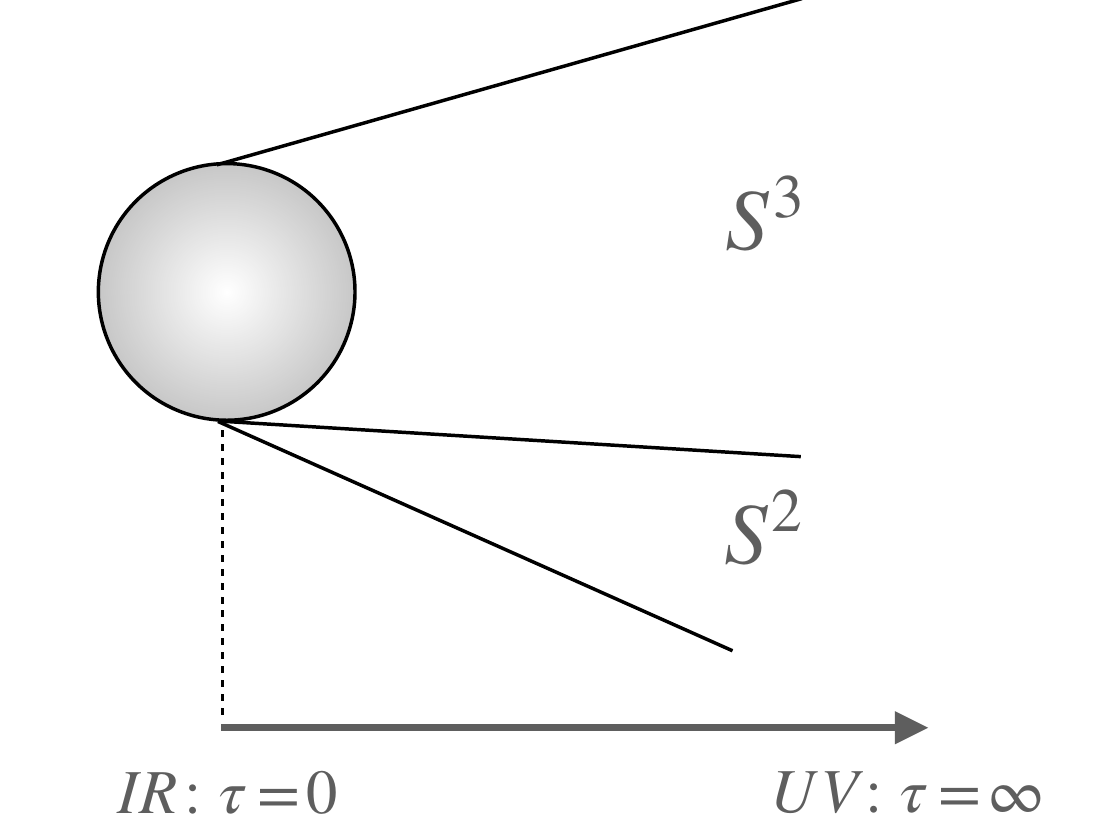}
\caption{An artist's impression of the KS geometry.}
\end{figure}
\vspace*{.2cm}

The most general deformation of the conifold metric with fluxes preserving the SU(2)$^2\times\mathbb{Z}_2$ symmetry can be written in terms of eight functions of a radial coordinate $\{\Phi_i(r)\}$ \cite{Papadopoulos:2000gj}; this space of type-IIB supergravity solutions includes the Klebanov-Strassler \cite{Klebanov:2000hb}, Maldacena-Nu\~nez \cite{Maldacena:2000yy}, and baryonic branch solutions \cite{Butti:2004pk}. In this Letter, we are interested in the deformation of the KS solution caused by the addition of $\overline{N}$ anti-D3 branes at the tip of the throat. In particular, we calculate how the anti-D3 branes affect the complexified three-form flux $G_3$, whose (p,q) components can be put in correspondence with various quantities in the effective four-dimensional low-energy theory describing the system.

Assuming that the backreaction of the anti-D3 branes on the geometry is small and can be studied in perturbation theory, the deformed geometry is given by
\begin{equation}
\Phi_i\,=\,\Phi_i^{KS}+\lambda\, \phi_i+O(\lambda^2)\,,
\end{equation}
where the analytical dependence of the fluctuations $\phi_i$ has been computed in \cite{Bena:2009xk,Bena:2011hz,Bena:2011wh} and the small expansion parameter is:
\begin{equation} \label{lambda}
\lambda=\frac{\overline{N}}{g_sM^2}\,,
\end{equation}
where $M$ is the integral of the Ramond-Ramond $F_3$-flux on the $S^3$. Usually, the number of anti-D3 branes is taken to be $\overline{N}=1$, since configurations with multiple anti-D3 branes have a tachyon \cite{Bena:2014jaa}, but we will keep track of it for completeness.

In the KS solution, the complexified three-form flux, $G_3$, is $(2,1)$ with respect to the choice of complex structure picked by supersymmetry \cite{Grana:2000jj}. When the anti-D3 branes are added at the tip of throat, the three-form flux also gets corrections, $G_3\!=\! G_3^{\text{KS}}\!+\!G_3^{\overline{\text{D3}}}$ and, at the same time, the complex structure is rotated. This implies that in general $G_3$ is not of $(2,\! 1)$-type anymore (and neither is imaginary-self-dual (ISD)) but it develops all other components: for example, the $(0,\!3)$ component  is 
\begin{equation}
\begin{split}
\label{eq:G03}
&G^{\overline{\mathrm{D3}}}_{(0,3)}= \frac{8\pi^2\lambda}{g_s Z_0}\frac{\partial_\tau\varphi(\tau)}{\sinh^2(\tau)} \,  \bar\Omega_{\text{KS}}+O(\lambda^2)\,,\\
&\varphi=g_s\,\sinh(\tau)\,\phi_7+\cosh(\tfrac{\tau}{2})\,\phi_5-\sinh(\tfrac{\tau}{2})\,\phi_6\,,
\end{split}
\end{equation}
where $\Omega_{\text{KS}}$ is the $(3,\!0)$ form defined by the KS complex structure, $Z_0$ is the fixed value of the conifold modulus and $\phi_{5,6,7}$ are functions of the radius, whose UV and IR expansions are in \cite{Bena:2009xk}, and whose full analytic expression can be found in \cite{Bena:2011hz}. This component  of the three-form flux generates a non-vanishing  (on-shell) Gukov-Vafa-Witten (GVW) superpotential \cite{Gukov:1999ya} (expressed in Planck units):
\begin{equation}
W_{\aD}\,=\,\int G_3^{\overline{D3}}\wedge \Omega\,.
\end{equation}
This integral can be performed explicitly, giving
\begin{equation} \label{WantiD310d}
W_{\aD}\, = \,-0.87\,i\,\,\lambda\,M\,Z_0 +O(\lambda^2) \,,
\end{equation}
where the first-order term is obtained using $\Omega_{\mathrm{KS}}$.
The anti-D3-brane not only generates this flux component, but also imaginary-anti-self-dual (IASD) pieces. These generate F-terms for the axio-dilaton and conifold moduli, given by 
\begin{equation}
D_{ \tau} W=-\frac{i\,g_s}{2} \int G^{*\aD}_{3}\wedge\Omega \ , \quad
D_Z W=\int G_3^{\aD}\wedge \chi \, , 
\end{equation}
where $\chi$ is a $(2,1)$-form (whose first-order expression in $\lambda$ can be found in \cite{Klebanov:1998hh}). Such integrals can be numerically evaluated using the explicit form of $G_-^{\overline{\text{D3}}}$ given in \cite{Massai:2012jn}:
\begin{equation}
\label{eq:F10d}
\begin{split}
D_ZW &= -1.5\,i\,\lambda\,M +O(\lambda^2)\,, \\
 D_\tau W&= 0.6\,\lambda\,g_sM\,Z_0 +O(\lambda^2) \,.
 \end{split}
\end{equation}
This on-shell superpotential and F-terms, computed using the ten-dimensional solution, will be used in section \ref{sec:deSitter} to compute an effective potential for the K\"ahler modulus in a KKLT-like construction.
\vspace*{-.5cm}
\section{4d supergravity description}
\label{sec:comparison}
\vspace*{-.3cm}
Before adding the $\aD$ branes at the tip of the throat, the superpotential and K\"ahler potential describing the conifold-modulus dynamics in a warped compactification has been computed in \cite{Blumenhagen:2019qcg,Dudas:2019pls}  :
\begin{equation}
\label{eq:W&K}
\begin{split}
W&=\frac{M}{2\pi i}\left( Z\log\!\frac{\Lambda^{3}_{\mathrm{UV}}}{Z}+Z+ w_Z\right)+i\frac{K}{g_s}Z\,,\\
\mathcal{K}&=-3\log\left(\rho+\bar\rho-\frac{\xi}{3}|Z|^{2/3}\right)+\log(2\gamma^4)
\,,
\end{split}
\end{equation}
where $\gamma^2=16\sqrt2 \pi^7 ||\Omega||^2$, $\xi=9c'g_sM^2$ and $c'\approx 1.18$ is a numerical factor coming from the warping \cite{Douglas:2007tu}.
Notice that the K\"ahler potential for the $Z$ modulus is known in a small-field expansion, and only the $Z^{2/3}$ term 
was worked-out explicitly. To avoid cumbersome expressions in what follows, we use the log form of the K\"ahler potential above (\ref{eq:W&K}), but it is understood that in the final results only the leading term in $Z^{2/3}$ is kept.

  The supersymmetric Minkowksi vacuum is given by:
 \begin{equation}
 \partial_Z W|_{Z_{KS}}\,=\,0\quad \Rightarrow\quad Z_{KS}=\Lambda_\text{UV}^3\,e^{-\frac{2\pi K}{g_s M}}\,. \label{ZKS}
 \end{equation}
Since the KS scalar potential and superpotential have to be zero on-shell in a supersymmetric Minkowski vacuum, this fixes the constant $w_Z$ in \eqref{eq:W&K}:
 \begin{equation}
 W_{\text{on-shell}}=0\quad \Rightarrow \quad w_Z \,=\,-\Lambda_\text{UV}^3\,e^{-\frac{2\pi K}{g_s M}}\,.  \label{ksw} 
 \end{equation}

We can promote this to an off-shell superpotential for the axion-dilaton as well, given by 
\begin{equation}
W_{\text{KS}}\,=\,\frac{M}{2\pi i}  \left[ Z\left(\log\frac{\Lambda_\text{UV}^3}{Z}+1\right) - \Lambda_\text{UV}^3 e^{\frac{2\pi i \tau K}{M}} \right] + K \tau Z \, .  \label{Wks}
\end{equation}
This satisfies  the supersymmetry condition in the axion-dilaton direction $D_\tau W|Z_{KS}=\partial_{\tau} W|_{Z_{KS}} = 0 $.

 We now add anti-D3 branes, whose backreaction can be captured in the language of the four-dimensional effective theory by:
\begin{itemize}
 
\item  an uplift term in the scalar potential,  breaking supersymmetry and shifting the conifold modulus vev from $Z_{KS}$ to $Z_0$ (to be computed below).
 
  \item A $(0,3)$ flux giving rise to an additional superpotential $W_{\overline{D3}}$, whose off-shell dependence on the conifold and dilaton-axion moduli will be determined by requiring  consistency with the ten-dimensional computation \eqref{WantiD310d}. 
 
\end{itemize}
 
To compute the former, it is useful to 
 describe the antibrane uplift potential in a manifestly supersymmetric way (more precisely in a non-linearly supersymmetric way) introducing a nilpotent chiral multiplet $X$ \cite{Antoniadis:2014oya,Ferrara:2014kva}, with the  following  K\"ahler potential and superpotential \cite{Dudas:2019pls} 
\begin{eqnarray}
\label{eq:KandW}
&&  {\cal K}  = - 3 \log \left( \rho + {\bar \rho} - \frac{|X|^2}{3} - \frac{\xi}{3} |Z|^{\frac{2}{3}} \right)  -  \log\left(\frac{\text{Im}\tau}{\gamma^4}\right) \ , \nonumber \\ 
 &&  W = W_{KS} + \frac{1}{M} \sqrt{ \frac{c''\,\overline{N}}{\pi} } Z^{2/3} \tau X   + A e^{- a \rho} +  W_{\overline{D3}} \,,  \ \label{sugra4d1}
\end{eqnarray}
where $c''\approx 1.75$ is a numerical factor related to the anti D3 brane energy \cite{Bena:2018fqc} and we have also included the non-perturbative contribution to the superpotential coming from gaugino condensation or D3-brane instantons.  
The four-dimensional scalar potential can then be written in the convenient form
\begin{eqnarray}
&& V =   \frac{\gamma^4g_s}{ r^2} \left\{ \frac{9}{\xi} |Z|^{4/3} |\partial_Z W|^2 +   |\partial_X W|^2 +  \frac{4}{g_s^2 r} | D_{\tau} W|^2 \right. \nonumber \\
&& \left.  +  \frac{\rho + {\bar \rho}}{3} |\partial_{\rho} W_{\text{eff}} -  \frac{3}{\rho + {\bar \rho}} W_{\text{eff}}|^2 -   \frac{3}{\rho + {\bar \rho}} |W_{\text{eff}}|^2    \right\} \ ,  \ \label{sugra4d2}
\end{eqnarray}
where 
\begin{equation}
W_{\text{eff}} = \frac{M}{2\pi i} \left( Z -\Lambda_\text{UV}^3 e^{- \frac{2\pi K}{g_s M}} \right) + A e^{- a \rho} +  W_{\overline{D3}}  \  , \ \label{sugra4d3}
\end{equation}
and $ r \equiv \rho + {\bar \rho} - \frac{\xi}{3} |Z|^{2/3} $ and where we used the on-shell axion-dilaton value $\text{Im}\tau = \,g_s^{-1}$.  In the absence of the non-perturbative term (setting $A=0$), 
the second line in  (\ref{sugra4d2}) is zero and the modulus  $\rho$ becomes a flat direction of the scalar potential, the so-called no-scale modulus.   In this limit the scalar potential becomes
\begin{equation}
\nonumber
V  =   \frac{\gamma^4|Z|^{4/3}}{c' r^2} \left\{  |  \frac{1}{2\pi i} \log\frac{\Lambda_\text{UV}^3}{Z}  + \frac{i K}{g_s M} |^2 +   \frac{c'\,c''}{\pi}\lambda  \right\}  + \cdots  \ .  \ \label{sugra4d4}
\end{equation}
The term in brackets  is the KS + uplift scalar potential,  whereas the $\cdots$ denote terms of order $Z_0^2$ and are subleading. As a function of the conifold modulus $Z$, this potential has a minimum at
 \begin{equation}
 Z_0  \simeq \left( 1 - \frac{8 \pi c' c''}{3}\lambda  \right)  Z_{KS} \ \equiv Z_{KS} + \delta Z \ , \label{4d1}
  \end{equation}
where we expanded to the first order in the $\overline{D3}$ uplift parameter $\lambda$, defined in \eqref{lambda}, in order to compare to the 10d computation. 

The on-shell value of the superpotential is then computed to first order 
 \begin{eqnarray}
 \nonumber
&& W (Z_0) \simeq W_{KS} (Z_{KS} ) + \partial_Z W_{KS} (Z_{KS}) \delta Z + W_{\overline{D3}}   \\
&&  = W_{\overline{D3}} = -0.87\,i\,\lambda\,M Z_0 \ , \ \label{4d2}
 \end{eqnarray}
where in the first line we used  (\ref{ZKS}) and  (\ref{ksw}) and in the second line we inserted the 10d input \eqref{WantiD310d}. 

The F-term of the conifold modulus can similarly be evaluated
 \begin{eqnarray*}
&&D_Z W (Z_0) = \partial_Z W (Z_0) +  {\cal K}_Z   W (Z_0)\,\simeq \\
&&\partial_Z^2 W_{KS} (Z_{KS}) \delta Z+  {\cal K}_Z W_{\overline{D3}}   \simeq  \frac{4 c' c''}{3 i}\lambda\,M  \simeq  -2.75\,i\lambda\,M \,, \  \ \label{4d3}
 \end{eqnarray*}
where  in deriving the result we anticipated, using the explicit form of the K\"ahler potential in \eqref{sugra4d1},  that the term ${\cal K}_Z W_{\overline{D3}} \sim O(Z_0^{2/3})$ and is therefore
subleading. This F-term has the same parametric dependence as its 10d counterpart  (\ref{eq:F10d}), with a different numerical coefficient.  

Finally, the F-term of the axion-dilaton is 
\begin{eqnarray*}
&&D_{\tau} W (Z_0)  \simeq  \partial_Z \partial_{\tau} W (Z_{KS}) \delta Z+  \partial_{\tau} W_{\overline{D3}}  +  {\cal K}_{\tau} W_{\overline{D3}}  \nonumber \\
&& \qquad \qquad \,\,\,\,= -  \frac{8 \pi c' c'' K}{3}\lambda\,  Z_{KS} +  \frac{0.435}{M}  Z_0 +  \partial_{\tau} W_{\overline{D3}}  \ .  \ \label{4d4}
\end{eqnarray*}
In order to obtain the correct parametric dependence of the ten-dimensional result (\ref{eq:F10d}) we impose
\begin{equation}
\partial_{\tau} W_{\overline{D3}} =  \frac{8 \pi c' c'' K}{3}\lambda\,  Z_{KS} \simeq K (Z_0-Z_{KS}) \ , \ \label{4d5}
\end{equation}
 and thus   
 \begin{equation}
  D_{\tau} W (Z_0) =  0.435\,g_sM\lambda\,Z_0 \ . \ \label{4d6}
\end{equation}
 The off-shell value of  $W_{\overline{D3}}$ should therefore be considered as an expansion 
\begin{equation}
W_{\overline{D3}} (\tau) = W_{\overline{D3}} (\tau_0) +  \partial_{\tau} W_{\overline{D3}} (\tau_0) (\tau - \tau_0) + \cdots  \ , \ \label{4d7}
\end{equation}
where we have determined the first two coefficients $W_{\overline{D3}} (\tau_0)$ and $\partial_{\tau} W_{\overline{D3}} (\tau_0)$ by consistency with the 10d results.

Let us stress that we  do not expect exact numerical agreement between the ten and four-dimensional results, but we do get the same parametric dependence. One of the reasons that the numerical factors might not exactly match is that the four-dimensional theory misses the effects of massive but light modes of the compactification \cite{Blumenhagen:2019qcg}.

Before closing this section, note that
the complete scalar potential (\ref{sugra4d2}) has an approximately decoupled structure
\begin{equation}
V = V_{\text{KS}+\text{uplift}} + V_{\text{KKLT}}  \ ,  \ \label{sugra4d5}
\end{equation}
with
\begin{equation}
V_{\text{KKLT}} = \frac{\gamma^4g_s}{r^2}  \left\{   \frac{\rho + {\bar \rho}}{3} |\partial_{\rho} W_{\text{eff}} -  \frac{3 W_{\text{eff}}}{\rho + {\bar \rho}} |^2 -   \frac{3 |W_{\text{eff}}|^2}{\rho + {\bar \rho}}     \right\}  + V_{ \overline{D3}} \ \nonumber  \ 
\end{equation}
and 
\begin{equation}
V_{ \overline{D3}} = V_{\text{KS}+\text{uplift}} (Z_0) \simeq \frac{c''|Z_0|^{4/3}}{\pi } \frac{\gamma^4}{(\rho + {\bar \rho})^2}\,\lambda  \  .   \label{sugra4d7}
\end{equation}
Furthermore, the  KKLT small superpotential constant $W_0$ is given in our construction by the on-shell value of the $\rho$-independent term in (\ref{sugra4d3}) 
\begin{eqnarray}
W_{0,\text{KKLT}}  = && \frac{M}{2\pi i} \left( Z_0 -\Lambda_\text{UV}^3 e^{- \frac{ 2\pi K}{g_s M}} \right) +  W_{\overline{D3}}  \nonumber \\
&& \simeq -i\left(0.87 - \frac{4}{3} c'c''\right)\lambda\,M\,Z_{KS} \ .   \ \label{sugra4d8}
\end{eqnarray}


\section{Bare-Bones de Sitter}
\label{sec:deSitter}
\vspace*{-.3cm}
In this section, we show that for certain choices of the parameters, the potential:
\begin{equation}
V=e^{\mathcal{K}}\left(G^{i\,\bar{j}} D_iW \overline{D_j\!W}-3 |W|^2\right)
\end{equation}
leads to de Sitter vacua. The potential is computed using the 10d input $D_iW$, $W= W_{\overline{\text{D3}}}+A e^{-a\rho}$ and K\"ahler potential as in \eqref{eq:KandW}.

\begin{figure}[!ht]
\centering \includegraphics[scale=0.45]{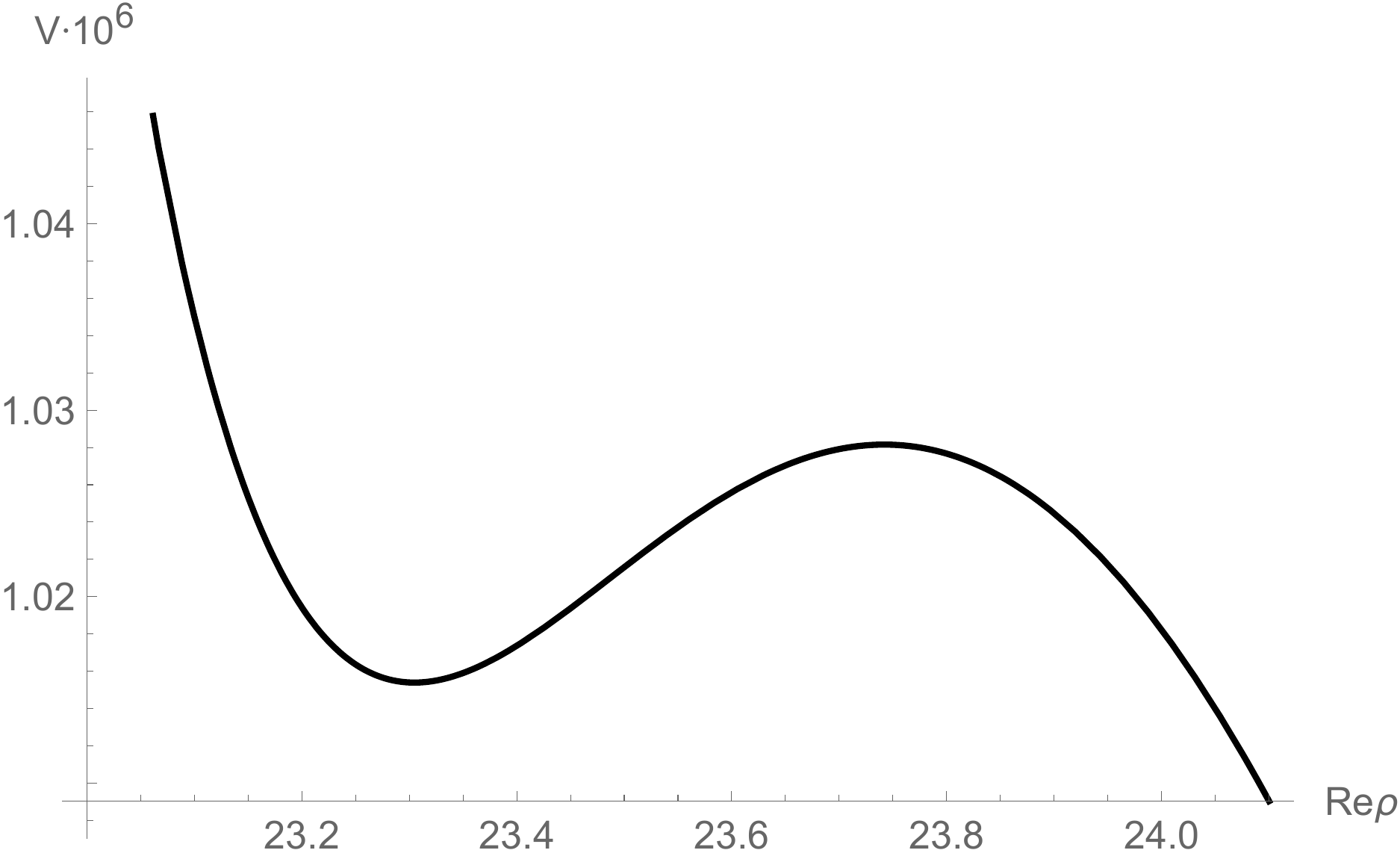}
\caption{The potential for the choice of parameters $a\!=\!\frac{\pi}{3}\,,g_s\!=\!\frac{1}{2}\,,A= 3\cdot 10^{3}\,, K=134\,, M=200$. This gives $Z_0\approx 10^{-4}$.}
 \label{Fig:deSitter}
\end{figure}

In Figure \ref{Fig:deSitter} we plot the potential for a particular choice of parameters.
This choice is not unique: we have performed a partial scan of the parameters and we are able to find several other de Sitter vacua.

For our de Sitter minimum the hierarchy between the bottom of the KS throat and the UV scale is of order $\frac{2\pi K}{g_sM}\approx 8$. For other de Sitter constructions without a large warping, see \cite{Bento:2021nbb}.

In the future, it would be important (but rather non-trivial) to check if the existence of this minimum survives higher order corrections in $\lambda$ and $Z$, as well as quantum corrections. 
\vspace*{-.5cm}
\section{Conclusions}
\vspace*{-.3cm}
A non-vanishing on-shell Gukov-Vafa-Witten superpotential is crucial in a KKLT-like construction of de Sitter vacua. In this Letter, we have shown that a small GVW superpotential, dubbed $W_{\overline{D3}}$ above, is generated by $\overline{D3}$ branes at the tip of a KS throat. This superpotential, together with the anti D3-brane-generated F-terms provide all that is needed to obtain a compactification with a positive cosmological constant. 

As we explained above, our proposal for constructing de Sitter solutions is more bare-bones and hence more robust than the KKLT one, because it has one less ingredient. Of course, as in all phenomenological constructions, adding more ingredients gives one more freedom to tune the resulting physical parameters. Hence, one can argue that our proposal, though more robust, is less accommodating that the KKLT construction for obtaining a parametrically small cosmological constant. However, the aim of our Letter is not phenomenological, bur rather to understand which ingredients are absolutely necessary to construct de Sitter, and which are optional, with an ultimate purpose of achieving a robust construction that may provide a way to escape the no-go arguments of \cite{Obied:2018sgi}. We believe our result represents a step in that direction.

Another interesting result of the calculation presented in this Letter is the parametric agreement between the first-principle, ten-dimensional computation of the effective potential (in section \ref{sec:fluxes}) and the four-dimensional-supergravity computation (in section \ref{sec:comparison}). To our knowledge, this is the first confirmation of the validity of the off-shell four-dimensional warped effective action \cite{Douglas:2007tu} and the analysis of \cite{Bena:2018fqc}.

Last, but not least, our proposal does not avoid some of the known constraints on KKLT-like models. It would be interesting to explore whether the problems underlined in \cite{Gao:2020xqh} also apply to our model. Furthermore, the minimum we found requires the contribution to the D3 tadpole of the fluxes in the KS throat to be of order $KM\approx 2\cdot 10^4$. In \cite{Bena:2020xrh} it was conjectured that such throats cannot be embedded in a flux compactification with stabilized moduli. It would be interesting to use our procedure to search for vacua where this tadpole contribution is smaller.

\noindent
{\bf Acknowledgments} We would like to thank Severin L\"ust for useful discussions.
This work was supported in part by the ERC Grants 772408 ``Stringlandscape'' and 787320 ``QBH Structure'', by the John Templeton Foundation grant 61149 and the ANR grant Black-dS-String  ANR-16-CE31-0004-01.

\bibliography{Draft}

\end{document}